# Reconciled warning signals in observations and models imply approaching AMOC tipping point

Yechul Shin[1], Ji-Hoon Oh[2], Sebastian Bathiany[3,4], Maya Ben-Yami[3,4], Marius Årthun[5,6], Huiji Lee[1], Tomoki Iwakiri[1], Geon-Il Kim[1], Hanjun Kim[7], Niklas Boers[3,4,8]*, and Jong-Seong Kug[1]*

[1]School of Earth and Environmental Sciences, Seoul National University, Seoul, Republic of Korea

[2]Scripps Institution of Oceanography, University of California San Diego, La Jolla, USA.

[3]Earth System Modelling, School of Engineering and Design, Technical University Munich, Munich, Germany

[4]Potsdam Institute for Climate Impact Research, Potsdam, Germany

[5]Geophysical Institute, University of Bergen, Bergen, Norway

[6]Bjerknes Centre for Climate Research, Bergen, Norway

[7]Department of Earth and Atmospheric Sciences, Cornell University, Ithaca, USA

[8]Department of Mathematics and Global Systems Institute, University of Exeter, Exeter, UK

*Corresponding author: Niklas Boers (boers@pik-potsdam.de) and Jong-Seong Kug (jskug@snu.ac.kr)

**Abstract**

Paleoclimate proxy records and models suggest that the Atlantic Meridional Overturning Circulation (AMOC) can transition abruptly between a strong and a weak state. Empirical warning signals in observational fingerprints indeed suggest a stability decline and raise concerns that the system may be approaching a tipping point. However, state-of-the-art Earth System Models (ESMs) do not consistently show such a stability loss, hence inconclusive in their projections of AMOC collapse under global warming. It remains unclear whether warning signals of AMOC tipping are overlooked in ESMs or overinterpreted in observations, calling for further investigations of AMOC stability. Here, based on the concept of critical slowing down, AMOC stability decline that can be interpreted as a warning signal of AMOC tipping is identified in the eastern SPNA in both observations and ESMs. This warning signal is in accordance with a physical indicator of AMOC stability—AMOC-induced freshwater convergence into the Atlantic basin. The observed signal can be reconciled with the modeled one only under warming exceeding the Paris Agreement goal, suggesting that AMOC stability is overestimated in ESMs. Our results suggest that the observed AMOC may indeed be losing stability and could thus reach a tipping point much earlier than state-of-the-art ESMs suggest.

**Introduction**

Carbon dioxide levels in the atmosphere have been steadily rising over the past century due to anthropogenic emissions. This has led to a continuous rise of global mean temperature and sea level, a trend that will continue at least for the coming decades, depending on global mitigation efforts (Matthews et al. 2009; Allen et al. 2009). These gradual responses to the radiative forcing demonstrate why a linear perspective is typically employed to understand climate change and its impacts (Matthews et al. 2009). However, gradual forcing does not always guarantee a gradual response. The dynamics of the Earth system can also be strongly nonlinear, and it has been suggested that several components of the Earth system, called tipping elements, may possess multiple stable states. At certain thresholds, also called "tipping points", the state of such a system can shift dramatically in response to even small perturbations, and transitions between these states can lead to abrupt responses. The risk of reaching climate tipping points is expected to increase with ongoing anthropogenic climate and land-use changes (Lenton et al. 2008; Armstrong McKay et al. 2022; Boers et al. 2022; Wang et al. 2023). If triggered, the changes in the corresponding tipping elements pose a risk not only through abrupt changes but also through the potential irreversibility of the transition to an alternative stable state, at least on human timescales. Thus, the consequences of large-scale climate tipping events on ecosystems and society would be severe, and the uncertainties in the critical forcing levels, e.g. in terms of global mean temperature, remain substantial.

The Atlantic Meridional Overturning Circulation (AMOC), a system of ocean currents in the Atlantic Ocean moving warm waters northwards in the upper ocean and cold waters southward at depth, is one of the proposed tipping elements in the Earth system. Several abrupt global changes recorded in paleoclimate proxies, including the Dansgaard-Oeschger events, have been attributed to abrupt transitions of the AMOC and used as evidence for its tipping behavior and bistability (Brovkin et al. 2021; Henry et al. 2016). The AMOC's bistability is also consistently supported by a hierarchy of climate models, including conceptual and theoretical box models (Stommel 1961; Vettoretti et al. 2022), Earth system models of intermediate complexity (EMICs) (Rahmstorf et al. 2005), low-resolution general circulation models (GCMs) (Hawkins et al. 2011; Boulton et al. 2014; Klus et al. 2019), and some state-of-the-art Earth System Models (ESMs) (Liu et al. 2017; Van Westen et al. 2024; Ben-Yami et al. 2024a). The combined evidence points to a potential collapse of the AMOC under future anthropogenic forcing. This could cause significant global impacts, as the warm and salty upper branch of the AMOC contributes to moderating regional climate (Zhang et al. 2019) and to attenuating extreme weather events (Yin and Zhao 2021) around the Subpolar North Atlantic (SPNA). These warm water masses reach northern high latitudes, affecting sea-ice variability, marine ecosystems (Shin et al. 2024) and fisheries (Schmittner 2005), and coastal sea levels (Yin et al. 2009; Wang et al. 2024). The AMOC also plays a crucial role in cross-equatorial northward energy transport, which caused the tropical precipitation bands, which are crucial for the tropical monsoon systems and ecosystems, to be predominantly positioned north of the equator (Frierson et al. 2013; Kang et al. 2015). A substantial AMOC weakening, e.g. associated with an abrupt transition, would lead to a southward shift of the tropical precipitation bands, posing severe ecological and socioeconomic risks that are unprecedented in human history (Drijfhout et al. 2015), especially given that the consequences would persist for the foreseeable future. This calls for a substantially improved understanding of AMOC stability and its potential for collapse.

The AMOC is expected to weaken during the twenty-first century in response to global warming, which is a general consensus among studies using state-of-the-art ESMs (Weijer et al. 2020). However, the probability of its collapse, and the position of a potential critical forcing threshold, remain debated, particularly concerning the likelihood of such a collapse to occur before 2100 (Ben-Yami et al. 2024b). Concerns of ongoing AMOC stability loss primarily arise from observations, while we note that significant weakening of the AMOC or of individual currents has yet to be directly observed (Volkov et al. 2024; Frajka-Williams et al. 2023). However, direct AMOC monitoring covers less than the last three decades. To extend the AMOC time series further back in time, several fingerprints of AMOC variability, for example based on sea-surface temperature (SST) and salinity patterns, have been proposed (Rahmstorf et al. 2015; Caesar et al. 2018), indicating that the AMOC has not been as weak as presently for at least a millennium (Caesar et al. 2021).

Observation-based fingerprints have also been utilized to investigate the stability of the AMOC (Boers 2021). When a dynamical system gradually approaches a bifurcation point, at which a critical transition would occur, its rate of recovery decreases, indicating a loss of stability. This phenomenon is referred to as critical slowing down (CSD). Measured, for example, in terms of rising autocorrelation and variance (Scheffer et al. 2009), or more directly based on regression-based estimates of the recovery rate itself (Boers 2021; Smith et al. 2022), CSD has commonly been employed as an early-warning signal of abrupt changes (Dakos et al. 2008; Scheffer et al. 2009; Smith et al. 2022). Significant CSD has indeed been detected in various SST- and salinity-based AMOC fingerprints, even when the measurement uncertainties and effects of non-stationary observational coverage are thoroughly propagated (Boers 2021; Ben-Yami et al. 2023). We emphasize that these fingerprints are never perfect reconstructions (Kilbourne et al. 2022; Zhu and Cheng 2024), and CSD cannot give a perfect alarm (Wagner and Eisenman 2015; Zimmerman et al. 2025). Given the underlying uncertainties, stability indicators such as the ones based on CSD should not be extrapolated to infer a future tipping time (Ben-Yami et al. 2024b). Nevertheless, consistent identification of these statistics empirically raises concerns about the loss of AMOC stability and an approaching AMOC tipping point. The Intergovernmental Panel on Climate Change (IPCC) currently tends to support that there will not be an abrupt collapse of the AMOC before 2100 (Intergovernmental Panel On Climate Change 2023), since most global warming simulations have not shown such sudden shifts (Drijfhout et al. 2015). However, with either bias adjustment (Liu et al. 2017), parameter tuning (Peltier and Vettoretti 2014), or prolonged external forcing (Van Westen et al. 2024), the tipping behavior of the AMOC is still demonstrated in several ESMs (Jackson et al. 2023). Huge uncertainties in AMOC projection, despite consistent weakening trends, further deepen the concerns (Reintges et al. 2017). Therefore, it remains unclear whether the AMOC is too stable in state-of-the-art ESMs, or if the likelihood of an AMOC transition is smaller than suggested by the analysis of observational fingerprints. Given the severe consequences of a potential collapse, the discrepancy between observational data and ESM projections on the loss of AMOC stability urgently needs to be addressed.

Reconciling the warning signs detected in observations with the results of ESM projections is essential for accurately assessing the current stability and collapse risk of the AMOC. This study presents combined evidence from observations and ESM simulations, indicating that the AMOC may be approaching a tipping point faster than ESM projections suggest.

**Results**

**Consistent eastern SPNA CSD in the observations and CESM2**

The eastern SPNA fuels the lower limb of the AMOC through deep-water formation, and is hence a crucial region shaping this circulation (Petit et al. 2020). Recent observations of the overturning strength from the Subpolar North Atlantic Program (OSNAP) suggest that the eastern SPNA dominates the overall characteristics of the subpolar overturning, both in mean and variability, compared to the western SPNA (Lozier et al. 2019; Chafik and Rossby 2019). Ocean properties, salinity and temperature, in the eastern SPNA has been attributed to the AMOC changes (Robson et al. 2012; Bryden et al. 2020) and are thus included in the majority of AMOC fingerprints (Boers 2021). Recently, this region has experienced an unusually large freshening event (Fig. 1a-d) as a result of major changes in ocean circulation (Holliday et al. 2020; Asbjørnsen et al. 2021). Although several freshening events, known as Great Salinity Anomalies (GSAs), have occurred in the past, this event is unprecedented in both pattern and magnitude, being both localized in the eastern SPNA and exceeding magnitudes observed over the past 120 years (Holliday et al. 2020). The downstream propagation of this freshwater anomaly from the eastern SPNA is suspected to be a reason for a recent weakening of both the Nordic Seas convection (Almeida et al. 2023) and the Deep Western Boundary Current (DWBC) (Koman et al. 2024), which are key processes forming the North Atlantic Deep Water (NADW) and the lower branch of the AMOC. Such unprecedented fluctuations may indicate a stability decline and, under the assumption of bistable dynamics, an approaching tipping point in the system.

To elucidate potential stability changes in the eastern SPNA, we calculate the lag-1 autocorrelation (AR1), a metric widely used to identify CSD, of SST observations (Methods). SSTs are chosen to ensure an adequate timescale for measuring CSD. Note that previous studies adopted at least 50-year to 500-year windows to calculate AR1 for the AMOC (Boulton et al. 2014), but upper-ocean salinity observations at the necessary time scales are quite rare. SST observations provide both the advantage of longer observational records and smaller uncertainty (Supplementary Fig. 1). Four different observation-based SST datasets consistently indicate that the AR1 of SSTs in the eastern SPNA has been increasing during recent decades (Fig. 1f). Although the different observational datasets show a range of magnitudes, their temporal evolution is consistent. This is particularly true with the inclusion of the 1990s, which partly reflects improvements in observational accuracy and coverage. Furthermore, these results are neither sensitive to changing the size of sliding windows (Supplementary Fig. 2) nor instigated by autocorrelated noise (Boettner and Boers 2022) (Supplementary Fig. 3), corroborating the significance of the CSD signal that we detect in the eastern SPNA. Overall, observations provide compelling evidence of an ongoing stability decline in the eastern SPNA, which has been proposed to precede an abrupt climate change event that occurred before the industrial revolution, the Little Ice Age (Arellano-Nava et al. 2022). With further global warming at least in the coming decades, this destabilization is likely to continue, raising concerns about the potential collapse of the AMOC.

To address these concerns, it is essential to employ state-of-the-art ESMs, which may simulate CSD as realistically as possible (Fig. 1g-l). We utilize the large ensemble experiments (LENS2) conducted by the Community Earth System Model Version 2 (CESM2; Methods) (Rodgers et al. 2021). The experiments span from 1850 to 2014 in the historical experiments and continue from 2015 to 2100 in the Socioeconomic Pathways (SSP) 3-7.0 experiments. The large ensemble is initialized with varying AMOC states and weather perturbations, enabling the separation of internal variability from forced responses. This experimental design is thus highly suitable for measuring CSD, which relies on the forced response and could otherwise be biased by internal variability (Methods). As shown in Fig. 1g-h, the eastern SPNA shows a nonlinear response to greenhouse warming (Fig. 1g-h). The region exhibits a freshening of only 0.26 g kg$^{-1}$ until 2050. Thereafter, the freshening rate intensifies fivefold (Fig. 1g). SSTs in the model simulations also deviate from a gradual warming trend (Fig. 1h). In addition, consistent with observed signals, the AR1 of the eastern SPNA SSTs shows a substantial increase prior to these abrupt changes (Fig. 1i). The substantial response of the ensemble-averaged AR1 strongly supports that the variability of the system is altered by the forcing, which is consistent with the hypothesis that the system approaches a bifurcation point, where a critical transition is expected. The ensemble spread of both salinity and SSTs show a similar evolution to the AR1 (green lines in Fig. 1g-h). As the ensemble spread increases, the range of possible system states at any given time becomes broader. Hence, this expansion consistently points out the loss of stability in the ESM simulations. By decomposing signals into both time and frequency domains, wavelet analysis reveals that the observed sign of CSD is largely attributed to changes in decadal fluctuations (Supplementary Fig. 4a). To illustrate how climate patterns correspond to the observed CSD, we construct a composite of salinity anomalies at these time scales (Methods), which exhibit a similar spatial pattern with the observed GSA (Fig. 1j-l). Overall, such a similarity between the LENS2 simulations and the observations—in both temporal statistics and spatial patterns—suggests that analogous underlying dynamics may be driving the loss of stability in both cases.

While this close resemblance supports the use of LENS2 to investigate eastern SPNA CSD and AMOC stability, there remains a considerable gap in reconciling the timing of CSD emergence (Fig. 2a). The CSD emergence is defined as the year when its 15-year linear trend shows the largest change (indicated by vertical lines in Fig. 1 and 2), revealing that CSD appears much later in the LENS2 simulations (2040s) than in observations (1990s). If the CSD observed in the eastern SPNA truly reflects a stability decline of the AMOC, its emergence should also align with AMOC-specific characteristics. Such alignment would reinforce the reliability of CSD in the SSTs of the eastern SPNA as an indicator for AMOC tipping. The present-day AMOC strength, defined as the maximum streamfunction at 26.5°N to match the location of the RAPID array, is well reproduced in the CESM2 simulations from the 1990s to the 2010s, so this does not explain the timing difference in CSD emergence (thick vs thin lines in Fig. 2b).

Another possible explanation for the different timing of CSD emergence in observations and simulations is a difference in the AMOC stability changes between the model and the real world. To quantify this, we employ the $\Delta M_{OV}$, which is defined as the meridional overturning component of freshwater convergence across the Atlantic basin (Methods). $\Delta M_{OV}$ is often adapted as an effective indicator for the physical AMOC stability, as it theoretically represents the basin-scale salt-advection feedback (Rahmstorf 1996a; Dijkstra 2007). When it is positive, freshwater converges into the Atlantic basin. The associated AMOC weakening diminishes the freshwater convergence, and the accumulation of salt within the basin compensates for the initial weakening. Conversely, when freshwater diverges from the Atlantic basin, the initial AMOC weakening is further reinforced by accumulation of freshwater, creating a self-sustaining feedback loop. Note that this theoretical concept has been confirmed by the bifurcation characteristics of AMOC in GCMs (Liu et al. 2017; Van Westen et al. 2024). Current ESMs are known to overestimate $\Delta M_{OV}$, leading to a too stable AMOC (Liu et al. 2017). Likewise, the LENS2 simulations also overestimate the $\Delta M_{OV}$ compared to observations (colors in Fig. 2b). In the 1990s, $\Delta M_{OV}$ from ocean reanalysis marks 0.1 Sv and the CSD begins to emerge in the SST observations, whereas the LENS2 records a substantially larger $\Delta M_{OV}$, 0.22 Sv, during the same period. As global warming weakens the AMOC, the $\Delta M_{OV}$ also continuously decreases, influenced by both the southern and northern boundaries of the Atlantic basin (Supplementary Fig. 5). Once the physical AMOC stability $\Delta M_{OV}$ of LENS2 is reduced as much as in the observations, the CSD emerges in the model simulations; in other words, the modelled $\Delta M_{OV}$ around 2050 is similar to $\Delta M_{OV}$ in the observations around the year 2000. When accounting for this time lag, $\Delta M_{OV}$ is statistically indistinguishable between simulations and observations (inner plot in Fig. 2b and Supplementary Fig. 5). The remarkable alignment demonstrates how the statistical, CSD-based stability indicator corresponds well with the physical measure of AMOC stability in both simulations and observations, supporting the view that CSD in the eastern SPNA SSTs effectively reflects a loss of stability of the AMOC. Therefore, the significant CSD in observations likely indicates a considerable stability loss of the AMOC, which is delayed in the simulations due to the model bias toward a too stable AMOC.

**Abrupt stratification increases in the Nordic Seas with loss of AMOC stability**

It is worth considering whether a reduction in the physical AMOC stability $\Delta M_{OV}$ can reliably indicate impending abrupt changes. Negative $\Delta M_{OV}$ indicates a state where the AMOC slowdown intensifies freshwater convergence. As global warming continuously weakens the AMOC, a significant freshening, i.e. a gain in buoyancy, must occur within the Atlantic basin around the $\Delta M_{OV}$ sign change. We therefore analyze the upper-ocean density in the Atlantic, which is a source of descending branch of the thermohaline circulation, to identify associated responses with $\Delta M_{OV}$. Prior to $\Delta M_{OV} = 0$, the upper-ocean density over the SPNA broadly decreases as AMOC weakens, with the maximum decrease in the Labrador Sea (Fig. 3b). This is expected because this region is known for deep convection and NADW formation. However, the Labrador Sea does not exhibit a substantial density response after $\Delta M_{OV}$ becomes negative (Fig. 3c), indicating that the Labrador Sea deep-water formation is not strongly related to $\Delta M_{OV}$. However, a distinct feature is found in the Nordic Seas, which is another critical region for deep convection (Asbjørnsen and Årthun 2023; Koman et al. 2024). In this region, the upper-ocean density responds minimally to the weakening of AMOC when the basin-scale salt-advection feedback is negative (i.e., $\Delta M_{OV} > 0$). After $\Delta M_{OV}$ changes sign, however, the Nordic Seas experience abrupt and dramatical density decreases; some regions show more than 6 times faster density changes than before (Fig. 3a,d). The abrupt density decrease extends beyond the upper-ocean and contributes to substantial buoyancy gains across the water column (Fig. 3e), clearly implying increased stratification and reduced dense-water formation. Note that this abrupt stratification is predominantly driven by salinity changes, with both the spatial pattern and the vertical profile confirming this (Supplementary Fig. 6), and does not occur in the historical period (Supplementary Fig. 7). Overall, there is an evident stratification in Nordic Seas in the LENS2 simulations, indicating a pronounced decrease in deep water formation that coincides with the decrease in physical AMOC stability discussed above.

Since the North Atlantic Current transports warm and salty water to both the eastern SPNA and the Nordic Seas, the bifurcation-related statistical behavior of the two regions is coupled. The Nordic Seas density therefore contains statistical characteristics indicative of an approaching bifurcation point. Its AR1 increases from the 2030s onwards (Fig. 3a), and this increase is attributed to increased decadal variability (Supplementary Fig. 4b), which is overall consistent with the warning signs from the eastern SPNA. Furthermore, the abrupt stratification increase in the Nordic Seas is concurrent with the eastern SPNA CSD, when the AMOC substantially loses stability (colors in Fig. 3e). Overall, the model changes that occur with substantial loss of AMOC stability (around 2050s in the simulations) are qualitatively similar to the observed changes around the 2000s—salinity fluctuations and CSD over the eastern SPNA, as well as weakened deep convection in the Nordic Seas (Almeida et al. 2023).

The meridional density difference between the northern and southern boundaries of the Atlantic Ocean is widely regarded as a key factor in establishing the AMOC. Not only do conceptual model use this difference to quantify the flow strength (Stommel 1961; Wood et al. 2019), but it also accounts for the intermodel diversity of AMOC (Reintges et al. 2017). At the southern boundary (34°S), the density decreases linearly in response to greenhouse gas forcing, and there is no statistical evidence of declining stability (Supplementary Fig. 8b,d). Abrupt stratification at the northern boundary is hence responsible for the reduction in meridional density difference and the emerging warning signals (Supplementary Fig. 8). Hence, its abrupt decreases could be indicative of AMOC destabilization, in line with a reduction in the physical AMOC stability indicator $\Delta M_{OV}$.

**Underestimation of proximity to a possible AMOC tipping point in simulations**

In the LENS2 simulations, a substantial loss of physical AMOC stability, as well as CSD in the eastern SPNA, accompanies the abrupt collapse of dense water formation in the SPNA, theoretically implying AMOC destabilization (Stommel 1961; Wood et al. 2019). Since the observed CSD is similar to that seen in the model, our results strengthen the conclusion that the real-world AMOC has been losing stability in recent years. Indeed, an abrupt collapse, akin to the modeled simulation, could occur in reality, if we assume that CESM2 adequately represents critical features of the AMOC. To further test this hypothesis, we compare the CSD response to the physical AMOC stability changes in both LENS2 simulations and observations. The physical-statistical stability relationship is highly consistent between the observation and the model (Fig. 4a). Although the model's AR1 is slightly higher than the observations' AR1s, which may be due to its coarse resolution or crude physical processes, the overall trend is closely aligned in the $\Delta M_{OV}$-AR1 space, at least within the available observational period. In the observational data, AR1 increases as the $\Delta M_{OV}$ decreases (colors in Fig. 4a) when climatological $\Delta M_{OV}$ is less than about 0.1 Sv, which is consistently shown in the model simulations (black in Fig. 4a). The alignment indicates that the simulations and observations share a similar critical behavior of the AMOC, allowing us to utilize the model as a diagnostic of the current state of the AMOC. Note that the relationship between physical AMOC stability and statistical stability is statistically consistent across different model versions of CESM (Supplementary Fig. 9), which partly relieves concerns about different model configurations. We observe that $\Delta M_{OV}$ has been continuously decreasing, with its recent values nearly reaching zero—when the model simulates an abrupt collapse in the Nordic Seas convection. Our results thus present combined evidence from simulations and observations, supporting that the observed AMOC is losing stability and thereby increasing the probability of an abrupt collapse.

The observed and modeled CSD-based warning signals are reconciled with the help of the physical AMOC stability $\Delta M_{OV}$. This implies that neither thresholds in global-mean temperature nor attempts to estimate tipping times should be the sole criteria when assessing tipping elements in the climate system—contrary to many recent assessments (Armstrong McKay et al. 2022). Despite the general consistency in AMOC strength and global-mean temperature, $\Delta M_{OV}$ exhibits a significant bias between observations and LENS2, implying that the AMOC is too stably simulated in the model. Consequently, if we set the control parameter as the global-mean SST, the stability response no longer agrees between the observations and models (Fig. 4b). The observed CSD is as severe as that simulated in LENS2 only after the global-mean SST has risen by 1.5 K, far exceeding the Paris Agreement target (Fig. 4b). This clearly indicates that CESM2 yields a strongly delayed CSD, leading to underestimation of AMOC tipping risk and reducing opportunities to detect early-warning signals. We note that not only CESM2 but in fact most climate models are known to simulate an AMOC that is likely too stable compared to reality(Liu et al. 2017, 2014). Therefore, in terms of the discrepancy in AMOC stability, whether the early-warning signals are overestimated in observations or overlooked in models, our findings suggest the possibility of overlooked warning in biased ESMs, and hence support the severe loss of stability inferred from observations, indicative of an approaching AMOC tipping point in the real world (Boers 2021; Ben-Yami et al. 2023; Ditlevsen and Ditlevsen 2023).

Our study evaluates the current state of the AMOC to understand whether there is an ongoing loss of stability that might lead to a collapse of the AMOC if it exhibits alternative stable states. We identify the combined evidence for CSD in terms of rising autocorrelation, which is not only captured at the same location but also at the same time when the physical AMOC stability indicator $\Delta M_{OV}$ reduces significantly. As in other attempts (Ben-Yami et al. 2024b), a concise and reliable estimate of a possible tipping time is not possible based on our results, while it still supports an approaching AMOC tipping point. Even if the tipping point of a given system has already been passed, timely reversal of the forcing–i.e., reducing global mean temperatures–may still prevent a full transition of the system (Ritchie et al. 2021; Bochow et al. 2023). This indicates that there remains a window of opportunity even if an AMOC tipping point were imminent. Nevertheless, our combined analysis of observed and modeled CSD in combination with physical AMOC stability suggests that the observed statistics, indicating loss of stability, are already as severe as the modeled statistics associate with warming levels that strongly exceed the Paris Agreement target, calling for urgent and substantially increased climate mitigation efforts.

## Methods

### CESM2 Large ensemble experiment

We utilize the Community Earth System Model 2 (CESM2) (Danabasoglu et al. 2020) Large ensemble (LENS2) (Rodgers et al. 2021). The CMIP6-era model is composed of physical components, including the atmosphere, ocean, land, and cryosphere, that are fully integrated with both land and ocean carbon cycles. It utilizes the Community Atmospheric Model Version 6 (CAM6) and Community Land Model Version 5 (CLM5); both are implemented with a spatial resolution of approximately 1° and have 32 vertical levels. It also includes the Parallel Ocean Program Version 2 (POP2) and the Community Ice CodE Version 5 (CICE5): a nominal horizontal resolution of 1° and includes 60 vertical levels. In addition, the Marine Biogeochemistry Library (MARBL) is used for the ocean carbon cycle.

The LENS2 provides 100 members of historical and SSP370 scenarios, which cover 1850-2100. The 100 members are further divided into two 50-member subsets: one uses the original CMIP6 biomass burning aerosol and the other uses a smoothed biomass burning. In this study, we employ the first subset alone. Both macro and micro perturbations are used to generate initial conditions for large ensemble members. These initial conditions correspond to various AMOC states including the maximum, decreasing, minimum, and increasing, allowing us to assess the impact and variability of AMOC under global warming scenario.

### Observations and Reanalysis data

Since we need sufficient temporal coverage to calculate the CSD, we only use the SST dataset which provides at least a century of data: the Hadley Centre Sea Ice and SST dataset version 1.1 (HadISST) (Rayner et al. 2003), Centennial in situ Observation-Based Estimates SST2 data (COBE2) (Hirahara et al. 2014), Extended Reconstructed Sea Surface Temperature version 5 dataset (ERSSTv5) (Huang et al. 2017), and EN4.2.2 (Good et al. 2013; Cheng et al. 2014; Gouretski and Cheng 2020). We use two different salinity datasets: In Situ Analysis System (ISAS) (Gaillard et al. 2016) and EN4.2.2 (Good et al. 2013; Cheng et al. 2014; Gouretski and Cheng 2020). In addition, we use ORAS5 (Zuo et al. 2019) to calculate $\Delta M_{OV}$ and utilize the AMOC streamfunction from different reanalysis (European Union-Copernicus Marine Service 2020).

### Critical Slowing Down and lag-1 Autocorrelation (AR1)

*Theoretical background*: The occurrence of CSD can be formally explained on the basis of dynamic system theory. Consider a typical system that is well described by deterministic function $f$ and additive stochastic noise:

$$dx = f(x, r(t), t)dt + \sigma dW_t, \qquad (1)$$

where $x$ is the system state, $r$ is the control parameter, $dW_t$ denotes the increments of a Wiener process $W_t$, scaled to standard deviation $\sigma$. If the system is close to a stable equilibrium, it is an acceptable assumption to consider the locally linearized dynamics

$$dx \approx f'(x^*, r(t), t)(x - x^*)dt + \sigma dW_t, \qquad (2)$$

and $f(x^*) = 0$ by definition. Furthermore, let's assume that the control parameter changes sufficiently slowly, which satisfies the system can be continuously follow the quasi-equilibrium, the linear approximation can simply the Eq. (2):

$$\dot{x} \approx -\lambda(t)x + \sigma dW_t, \tag{3}$$

with $\lambda = -f'(x^*, r(t), t)$. The discretization of the Ornstein-Uhlenbeck process allows us to relate the $\lambda$ to the variance $var(x_t) = \sigma^2/2\lambda$ and lag-1 autocorrelation $AR1 = e^{-\lambda}$. This framework suggests that the loss of stability is essential when the system approaches the bifurcation point, and it manifests as decrease in $\lambda$; thus, continuous increase in variance and autocorrelation is widely adapted to measure CSD, as a early-warning signal of abrupt transition (Boers and Rypdal 2021; Bochow and Boers 2023).

*AR1 in LENS2*: Because the CSD is based on linearized dynamics near the stable equilibrium, the AR1 is calculated in the residual time series which the forced response is removed. Thus, it is important to calculate CSD that appropriately removes the forced component in a time series. Previous studies employed different method to remove a forced response such as Gaussian smoothing (Dakos et al. 2008) and Seasonal and Trend decomposition using Loess (STL method) (Boulton et al. 2022). Since these methods require parameters, such as window size, to remove the forced response, conducting sensitivity tests on these parameters is necessary. In this context, the large ensemble approach is particularly well-suited for measuring CSD, as the experimental design inherently provides robust estimations of both forced and internal processes. Furthermore, the large ensemble enables separation of the seasonal cycle, which provides an extensive dataset to detect changes in residual processes. In this study, we hence utilize monthly time series. We first remove the ensemble average for each ensemble member and then calculate a AR1 with 90-year window from the residual time series.

*AR1 in observations*: Although the large ensemble approach provides a credible way for removing the forced response, it cannot be directly applied to observations. Therefore, we adopt the large ensemble method as a benchmark to establish the most appropriate method for eliminating the forced response in observations. We begin by removing the nonlinear trend from the monthly time series by subtracting its moving average. Subsequently, we calculate the AR1 for each ensemble member. This procedure only utilizes information from individual ensemble members and does not require information about the ensemble mean. Next, we evaluate the impact of different window sizes to remove the forced response by comparing AR1s derived from individual members with those derived from the ensemble mean. Based on the correlation coefficient and root-mean-square error, we select a 44-year window (Supplementary Fig. 10a). The AR1 values obtained using the moving average demonstrate a linear relationship with those calculated from the ensemble mean (Supplementary Fig. 10b), thereby alleviating concerns regarding methodological dependency and validating the observation-based AR1 estimates.

**Composite analysis on SPNA salinity variability**

Both CSD and ensemble spread suggest that global warming intensifies upper-ocean variability in the eastern SPNA in the LENS2 (Fig. 1g-i). To visualize the spatial pattern of this enhanced variability, we conduct a composite analysis on the eastern SPNA. First, we apply a 5-to-20-year bandpass filter to the upper-ocean (0–200 m) salinity data to extract decadal fluctuations, as wavelet analysis indicates these time frequencies contribute to the CSD (Supplementary Fig. 11a). We then identify all freshening peaks in the eastern SPNA (17 total), defined as a local minimum where the freshening exceeds 3 historical standard deviation (Supplementary Fig. 11b). We collect these freshening peaks from 2030 to 2060, when the CSD and ensemble spread abruptly increase, and make a composite of salinity variability (Supplementary Fig. 11c). This composite depicts a spatial pattern that aligns with the CSD (Fig. 1j-l), and we compare this pattern with observed salinity fluctuations, which supports the similarity of critical behavior between the observed and ESM simulations in terms of spatial distribution.

**AMOC stability indicator**

Freshwater transport by the AMOC at the southern boundary of the Atlantic Ocean (~34°S) was proposed as a diagnostic indicator of AMOC stability (Rahmstorf 1996). Later, the northern boundary is also considered for freshwater transport by the AMOC, leading to the proposal of AMOC-induced freshwater convergence as an indicator of AMOC stability indicator (Dijkstra 2007), which has been validated in several climate models (Liu et al. 2017, 2014). The diagnostic indicator basically represents the basin-scale salt advection feedback, which has been pointed out as a fundamental cause of bistability (Stommel 1961). In this study, we also employ freshwater convergence to compare the ESM simulations with observations.

The AMOC freshwater transport at a certain latitude ($\phi$) can be calculated as

$$M_{ov}(\phi) = -\frac{1}{S_0} \int_{-D}^{0} \left[ \int_{W}^{E} v^* \, dx \right] [\langle S \rangle - S_0] dz \qquad (4)$$

where $v^* = v - \hat{v}$ is the baroclinic meridional ocean velocity, which subtracts $\hat{v}$ barotropic meridional ocean velocity, $\langle S \rangle$ indicates the zonal mean salinity, and $S_0 = 34.8$ g kg$^{-1}$ is a reference salinity. The AMOC stability indicator ($\Delta M_{ov}$) is defined as the difference of $M_{ov}$ across the southern and northern boundaries of the Atlantic: $\Delta M_{ov} = M_{ov,S} - M_{ov,N}$. We choose the southern boundary $M_{ov,S}$ at 34°S and the northern boundary $M_{ov,N}$ is consist of Canadian Arctic Archipelago, the Fram Strait, and the Barents Sea Opening (Liu et al. 2017, 2014). For northern boundaries, we follow the equation (4) with baroclinic ocean velocity normal to each section.

**Acknowledgments**

YS is supported by the National Research Foundation of Korea (NRF) grant funded by the Korea government (MSIT) (RS-2024-00334637; RS-2024-00438471). NB, SB, MB are funded by the Deutsche Forschungsgemeinscahft (DFG, German Research Foundation) – project number 551886647. This is ClimTip contribution #XX; the ClimTip project has received funding from the European Union's Horizon Europe research and innovation programme under grant agreement No. 101137601. NB acknowledges further funding by the Volkswagen foundation. MÅ is supported by The Research Council of Norway project Overturning circulation in the new Arctic (Grant 335255). HL, TI, GK, HK, JK are supported by the National Research Foundation of Korea (NRF) grant funded by the Korea government (MSIT) (RS-2022-NR070706).

**Figure list**

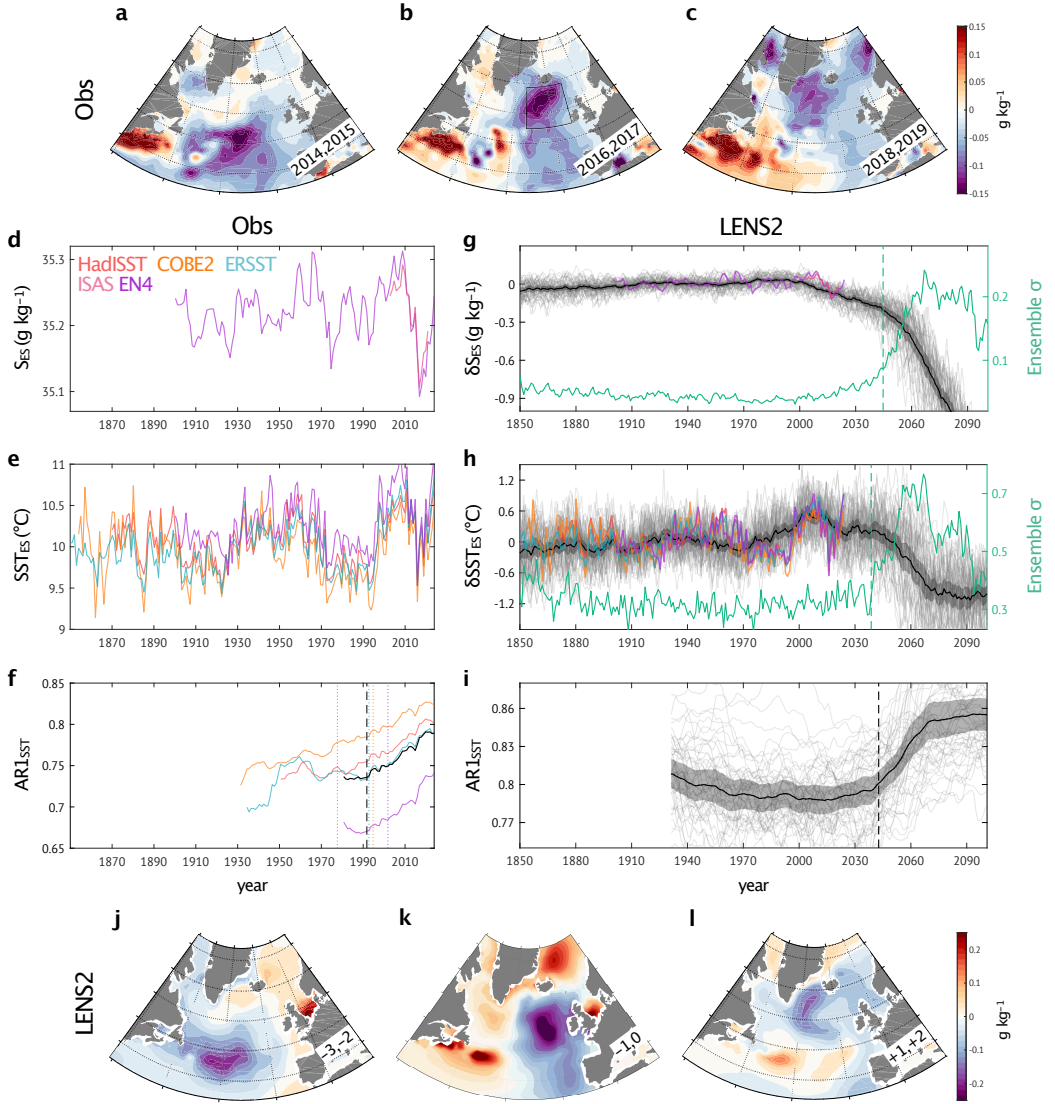

**Figure 1. Critical Slowing Down (CSD) in observations and ESM simulations. a-c**, Upper-ocean (0-200 m) salinity anomalies in the Subpolar North Atlantic (SPNA) for 2014-2015 (**a**), 2016-2017 (**b**), and 2018-2019 (**c**), relative to the 2004-2023 climatology. The black box in (**b**) marks the eastern SPNA. **d-e**, Time series of the eastern SPNA salinity (**d**) and sea surface temperature (SST) (**e**). **f**, Lag-1 autocorrelation (AR1) of the eastern SPNA SST (Methods). Each color represents corresponding observations, and the black line shows its average. **g-h**, Same as (**d-e**) but for the LENS2 simulations and anomalies relative to the 1900-1950 climatology. Achromatic lines represent LENS2: the thick line shows the ensemble mean, shading represents the 95% confidence interval, and thin lines indicate individual members. The green line represents interensemble spread, and colored lines correspond to observation anomalies. **i**, Same as (**f**) but for LENS2. The vertical line indicates where the metric (i.e., AR1) starts to increase, which is defined as the year in which a 15-year linear trend shows the largest trend. Dotted lines are for each observation, the dashed-dot line indicates its average, and the dashed line denotes the LENS2 ensemble mean. **j-l**, Upper-ocean salinity composite based on the eastern SPNA freshening event, shown before (**j**), around (**k**), and after (**l**) the freshening peak (Methods). Numbers represent the time lag in years from the freshening peak.

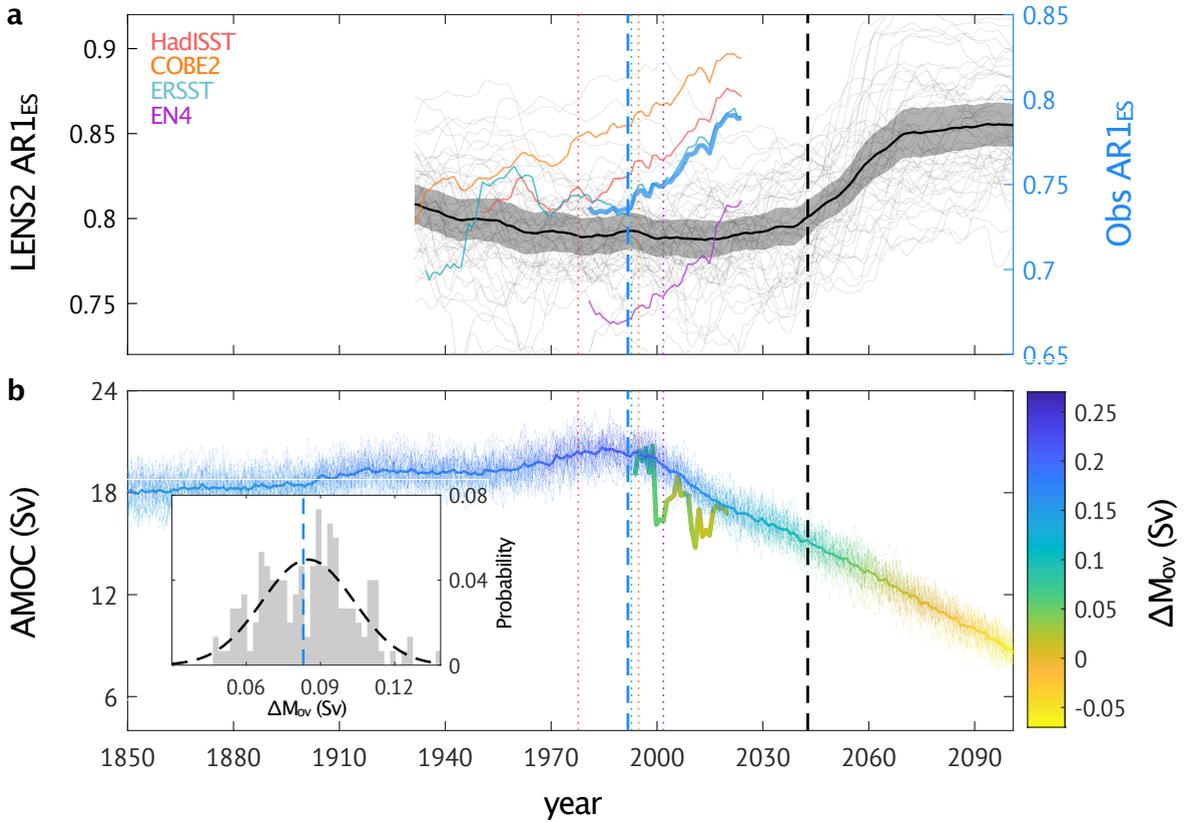

**Figure 2. AMOC stability and the CSD timing.** a, AR1 of eastern SPNA SST of observations (colors). Each color represents corresponding observations, and the thick blue line shows its average. The AR1 of LENS2 are shown in achromatic lines. The thick line shows the ensemble mean, shading represents the 95% confidence interval, and thin lines indicate individual members. b, Time series of AMOC strength from reanalysis datas (thick; Methods) and LENS2 (thin). The AMOC strength is defined as the maximum meridional streamfunction at the 26.5°N. Colors indicate $\Delta M_{OV}$, baroclinic freshwater convergence into the Atlantic basin, suggested as physical AMOC stability indicator (Methods). The CSD timing (vertical lines in Fig. 1f and 1g) are overlaid as a vertical line; blue dashed is for observations and dashed is for LENS2. The insert shows a histogram for LENS2 $\Delta M_{OV}$ spread at the CSD timing, a fitted normal distribution is shown by the dashed line. Corresponding $\Delta M_{OV}$ from reanalysis is shown by the dashed-dot line, indicating the $\Delta M_{OV}$ between the reanalysis and model simulations are statistically indistinguishable around the CSD timing.

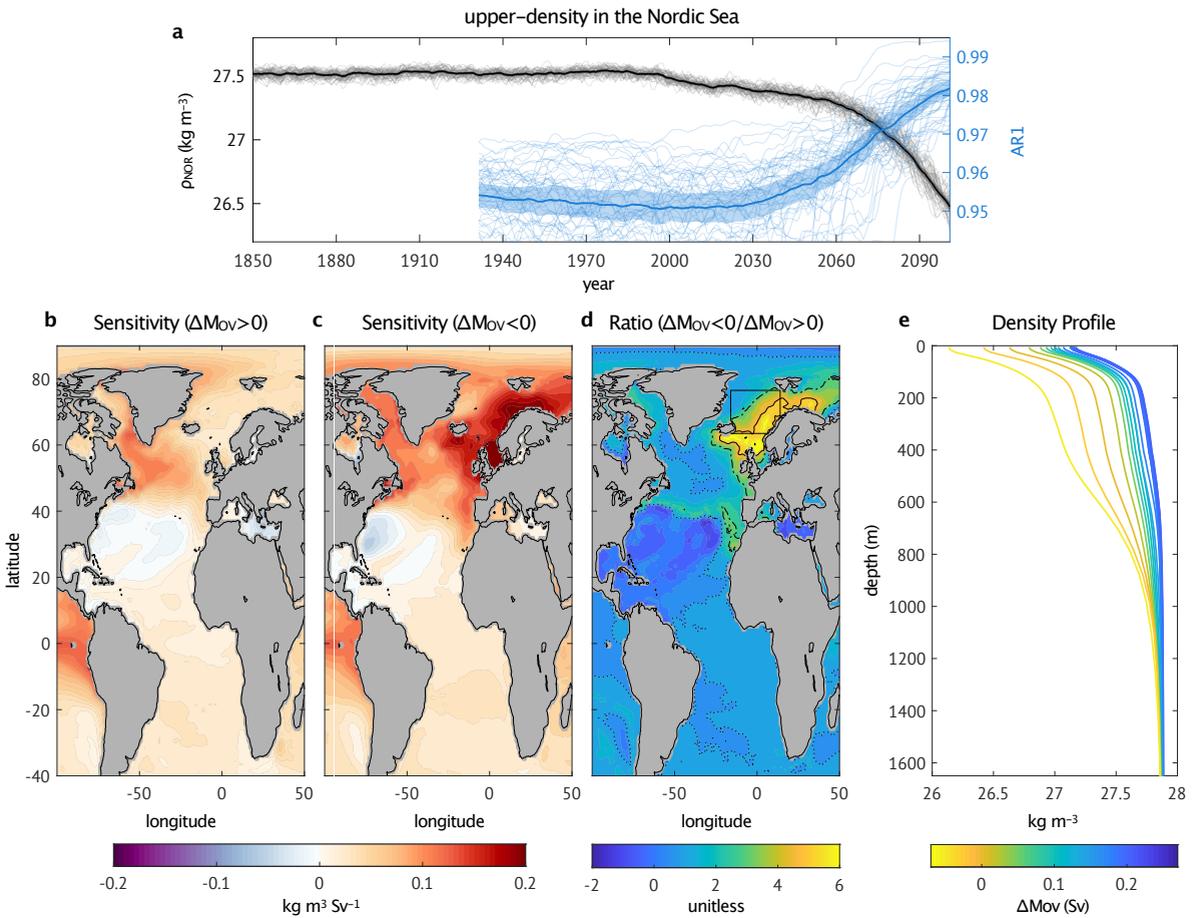

**Figure 3. Abrupt Nordic Seas stratification emerging with AMOC stability. a**, Time series of upper-ocean density at the Nordic Seas (black), marked as black box in (**d**), and its lag-1 autocorrelation (blue). Thick lines are ensemble mean, shading represents 95% confidence range, and thin lines are individual ensembles. **b-c**, Sensitivity of upper-ocean density to AMOC strength (kg m$^3$ Sv$^{-1}$) is measured by linear regression for 20 years before (**b**) and after (**c**) $\Delta M_{OV} = 0$. **d**, Ratio of sensitivities depicts an amplified relationship aligned with the sign of $\Delta M_{OV}$. Dotted, dashed-dotted, and solid black lines correspond to sensitivity ratios of 2, 4, and 6, respectively. Black box indicates the Nordic Seas. Both sensitivity and ratio are shown as ensemble medians. **e**, Vertical density profile evolution at the Nordic Seas. Each line indicates decade-averaged density profile, and its color is for $\Delta M_{OV}$, physical AMOC stability, same as Fig. 2b.

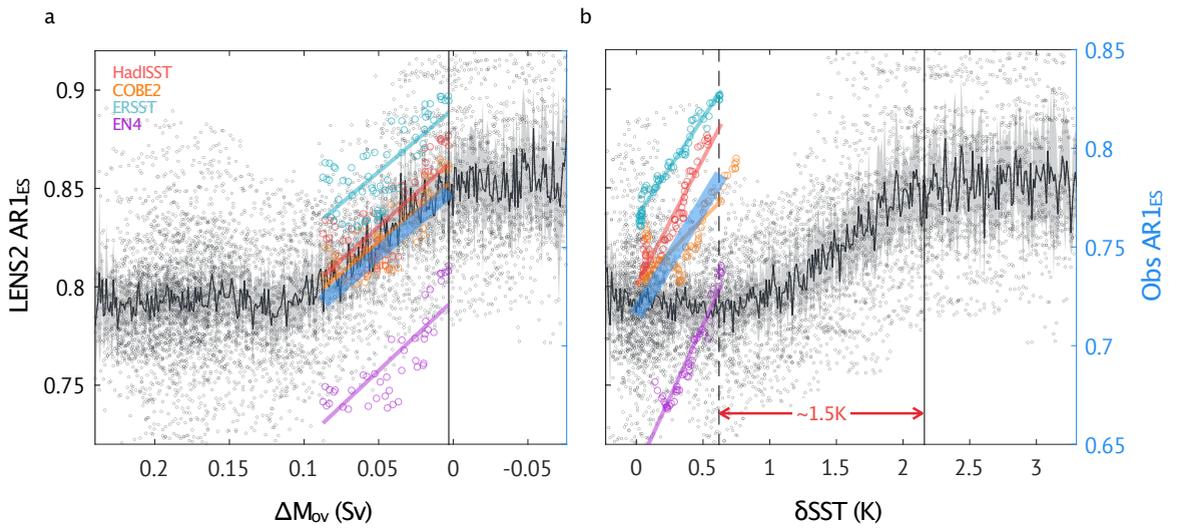

**Figure 4. Approaching tipping point in both observations and simulation. a**, Eastern SPNA SST AR1 as a function of $\Delta M_{OV}$ for LENS2 (left axis) and observation (right axis). The black solid line represents the mean, while gray shading indicates the 95% confidence interval. Observational data points are color-coded, linear fits are displayed as colored lines, and the blue line denotes the ensemble mean of these fits. The most recent $\Delta M_{OV}$ is marked by the vertical line (0.03 Sv). **b**, Same as (**a**) but as a function of global-mean SSTs relative to 1900. SST anomalies corresponding to the latest $\Delta M_{OV}$ are marked as vertical lines: dashed-dotted for observation (0.62 K) and dashed for LENS2 (2.16 K). Although there is a difference in magnitude, the $\Delta M_{OV}$-AR1 space is coherent for both observations and model simulations, suggesting recent CSD in the eastern SPNA is consistent what the LENS2 simulations. In contrast, the SST-AR1 relationship is mismatched due to model biases, with current CSD resembling the situation after 2 K warming in LENS2.

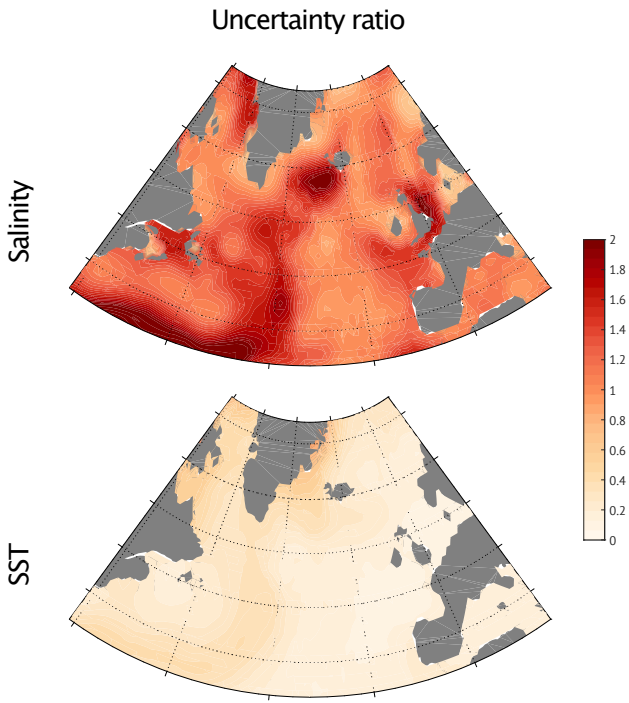

**Supplementary Fig. 1 SST and upper-ocean salinity over the SPNA.** Ratio of the standard deviation of the analysis error to the standard deviation of the upper-ocean (0-200 m) salinity and SST.

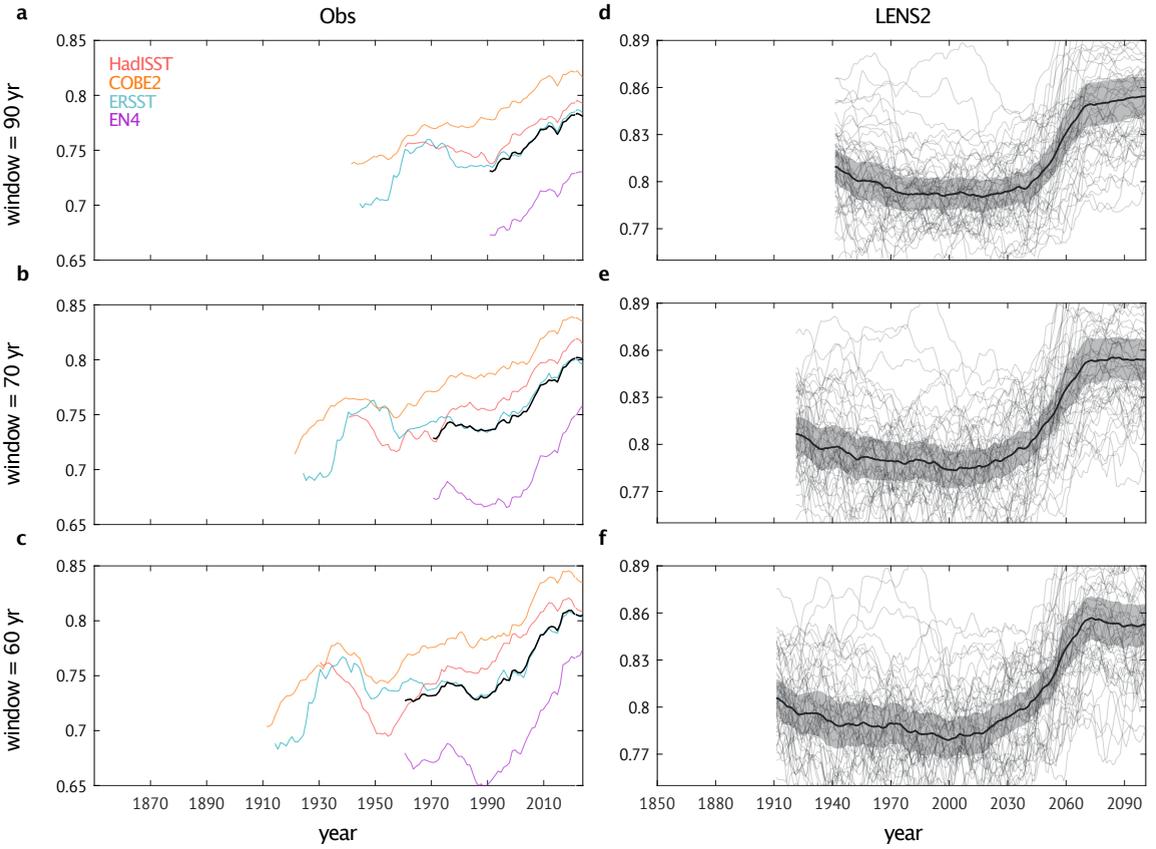

**Supplementary Fig. 2 Same as Fig. 1f and 1i but with varying sliding window size to estimate CSD. a-c**, Lag-1 autocorrelation (AR1) of eastern SPNA SST calculated by corresponding sliding window size. Each color represents corresponding observations, and the black line shows its average. **d-e**, Same as (**a-c**) but for LENS2. The thick line shows the ensemble mean, shading represents the 95% confidence interval, and thin lines indicate individual members.

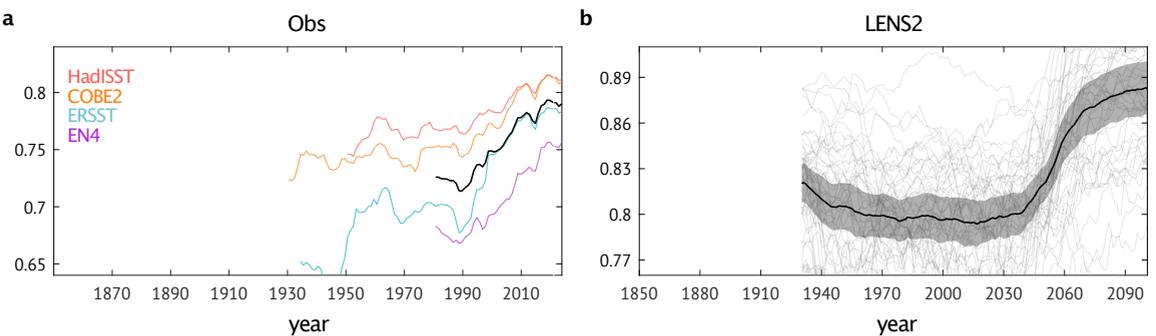

**Supplementary Fig. 3 Same as Fig. 1f and 1i but impact of autocorrelated noise is removed (Boettner and Boers, 2022). a**, Lag-1 autocorrelation (AR1) of eastern SPNA SST calculated by corresponding sliding window size. Each color represents corresponding observations, and the black line shows its average. **b**, Same as (**a**) but for LENS2. The thick line shows the ensemble mean, shading represents the 95% confidence interval, and thin lines indicate individual members.

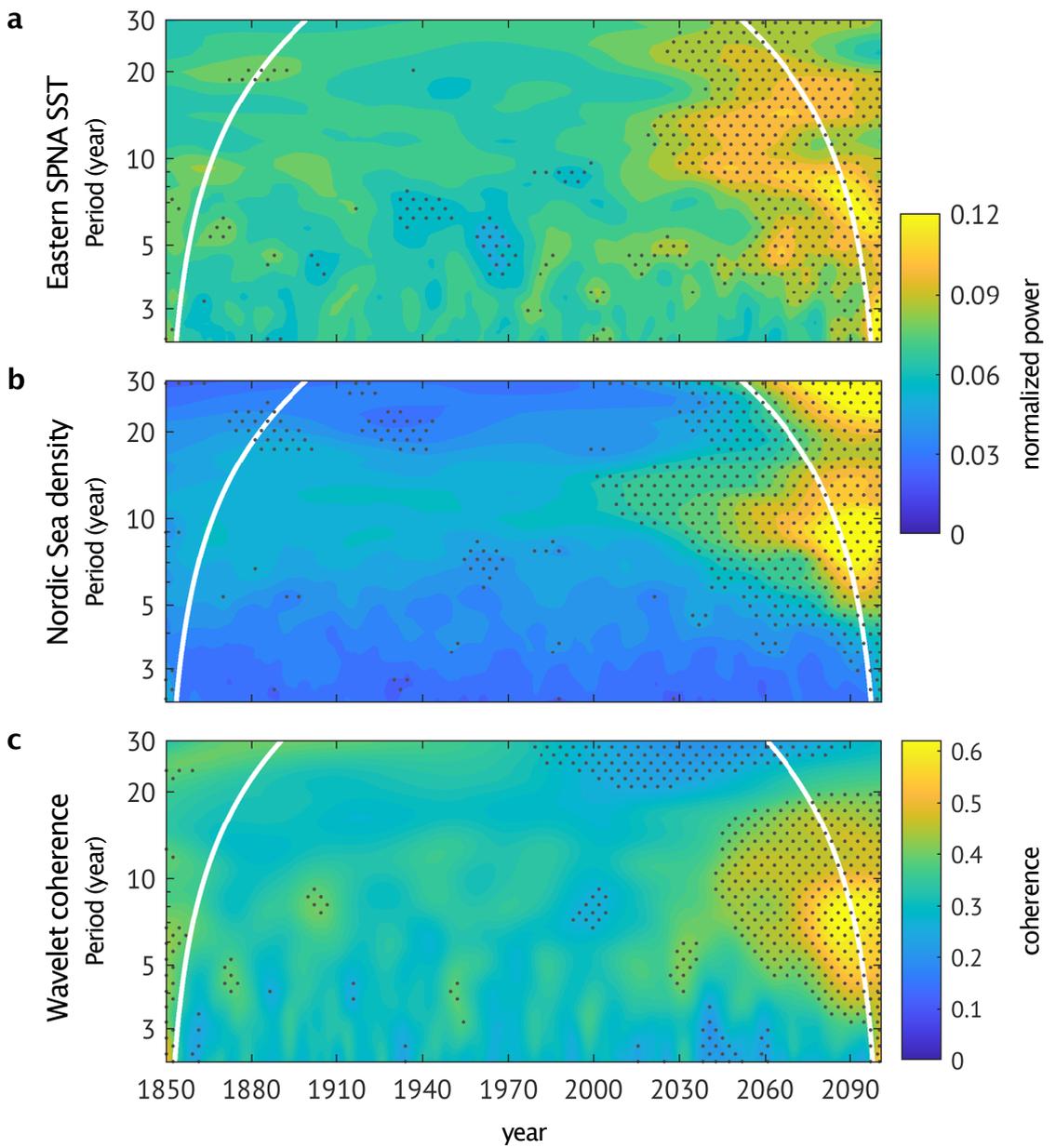

**Supplementary Fig. 4 Ensemble-averaged wavelet power spectrum. a-b,** Ensemble-averaged wavelet power spectrum of the eastern SPNA SST (**a**) and Nordic Sea density (**b**). **c**, Wavelet coherence between eastern SPNA SST and Nordic Sea density. The white line marks the cone of influence. Dotted areas indicate regions where the wavelet power is statistically significant at the 95% confidence level, as determined by bootstrap testing relative to the 1900–1950 baseline period.

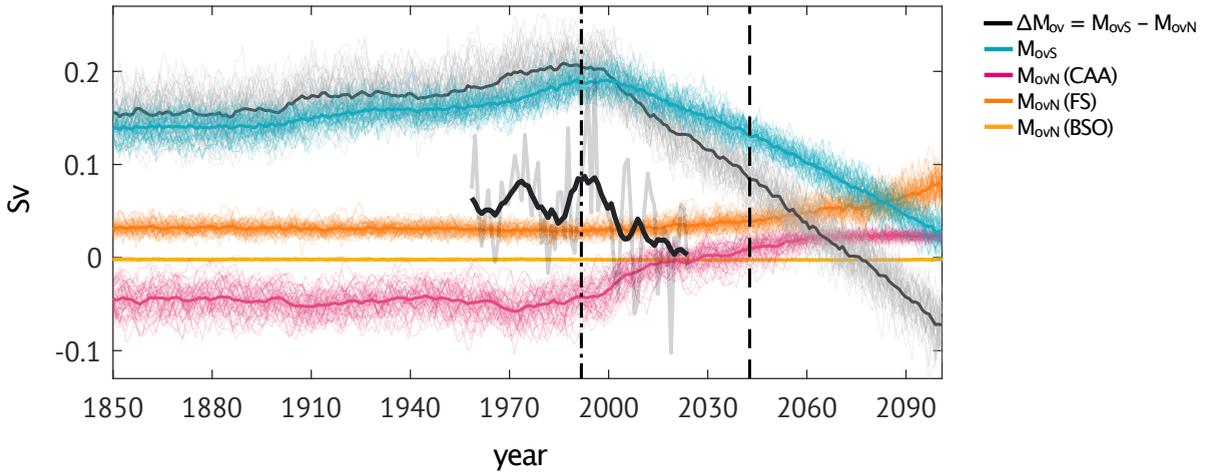

**Supplementary Fig. 5 AMOC stability time series in observations and ESM simulations.** Time series of annual-mean northward freshwater transport by meridional overturning circulation (Methods). Thick line denotes the Ocean Reanalysis System 5 (ORAS5), while thick black line is 11-year running average. Thin lines are from the LENS2. Vertical lines indicate the timing of CSD identified in Figs. 1 and 2, with dashed-dot lines corresponding to ORAS5 and dashed lines corresponding to LENS2.

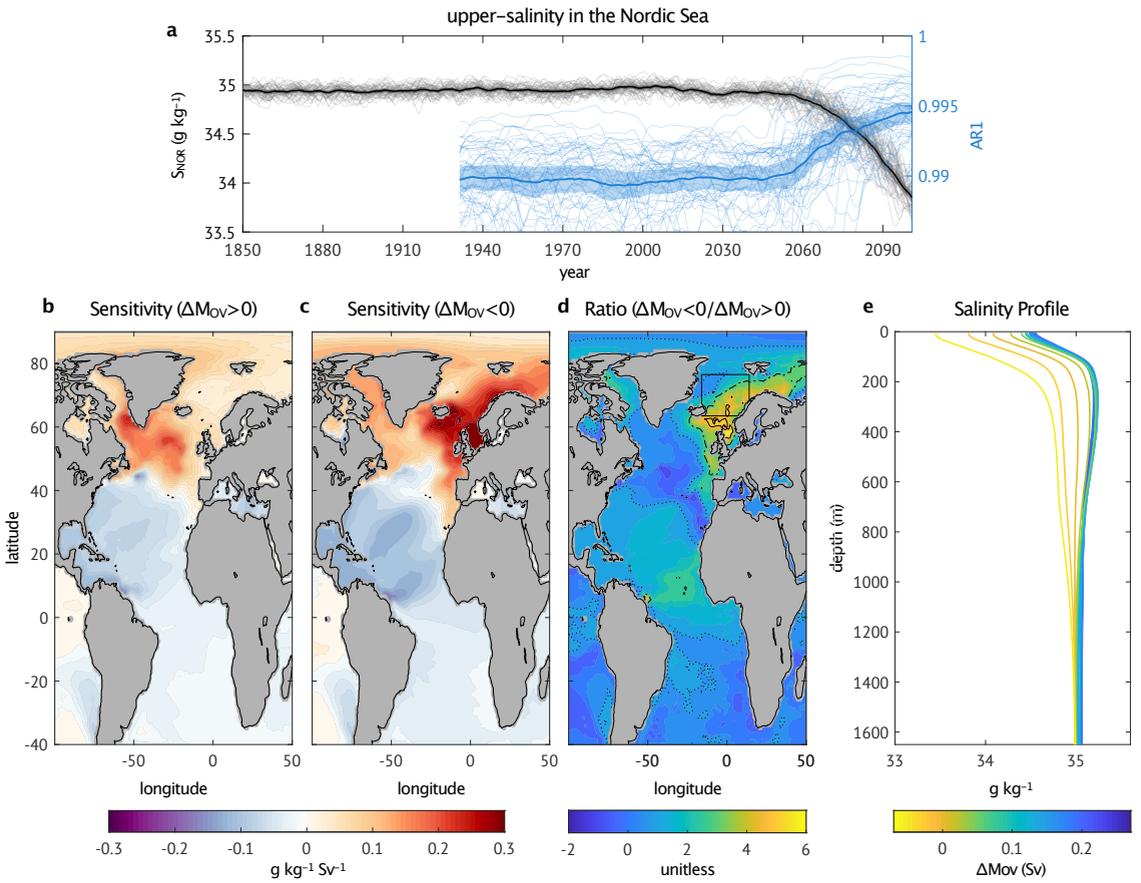

**Supplementary Fig. 6 Same as Fig. 3, but for upper-ocean salinity. a-b**, Sensitivity of upper-ocean salinity to AMOC strength (PSU Sv$^{-1}$) is measured by linear regression for 20 years before (**a**) and after (**b**) $\Delta M_{OV} = 0$. **c**, Ratio of sensitivities depicts an amplified relationship aligned with the sign of $\Delta M_{OV}$. Dotted, dashed-dotted, and solid black lines correspond to sensitivity ratios of 2, 4, and 6, respectively. Black box indicates the Nordic Sea. Both sensitivity and ratio are shown as ensemble medians. **e**, Vertical salinity profile evolution at the Nordic Sea. Each line indicates decade-averaged density profile, and its color is for $\Delta M_{OV}$, physical AMOC stability, same as Fig. 2b.

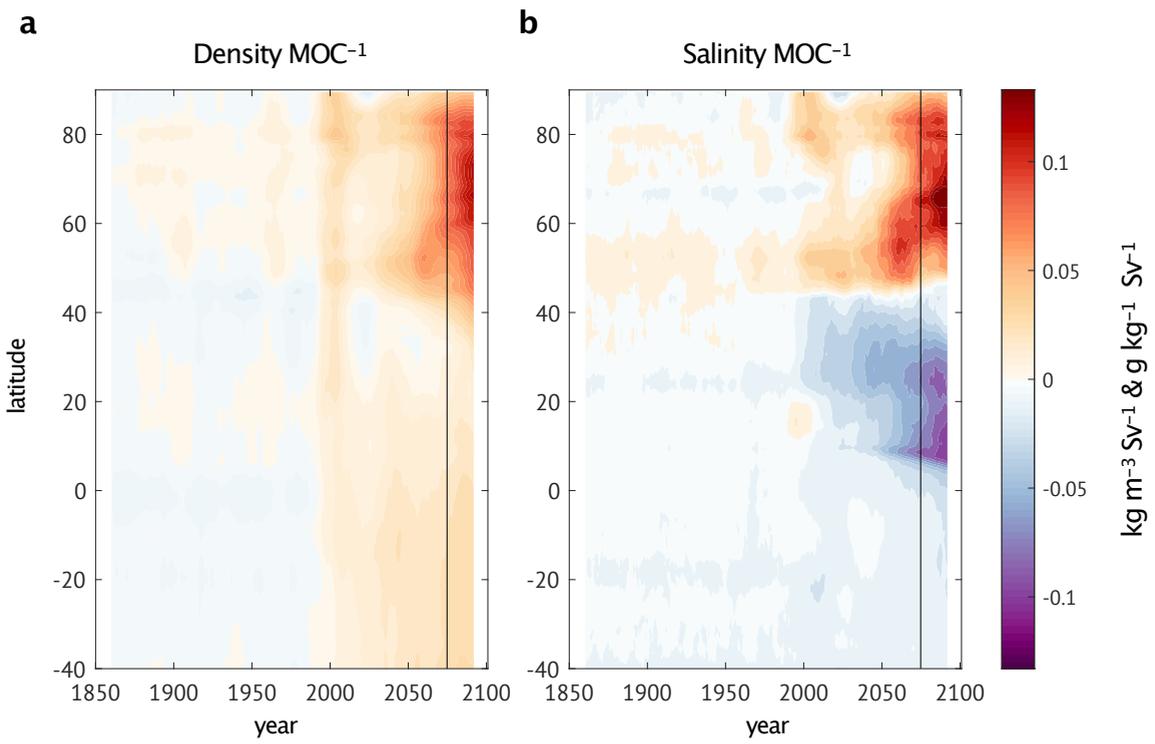

**Supplementary Fig. 7 Hovmöller diagram of Atlantic basin. a-b**, Zonal-mean sensitivity of upper-ocean (0–200 m) density (**a**) and salinity (**b**) to AMOC strength, derived using 20-year linear regressions. The vertical line denotes the time at which $\Delta M_{OV} = 0$

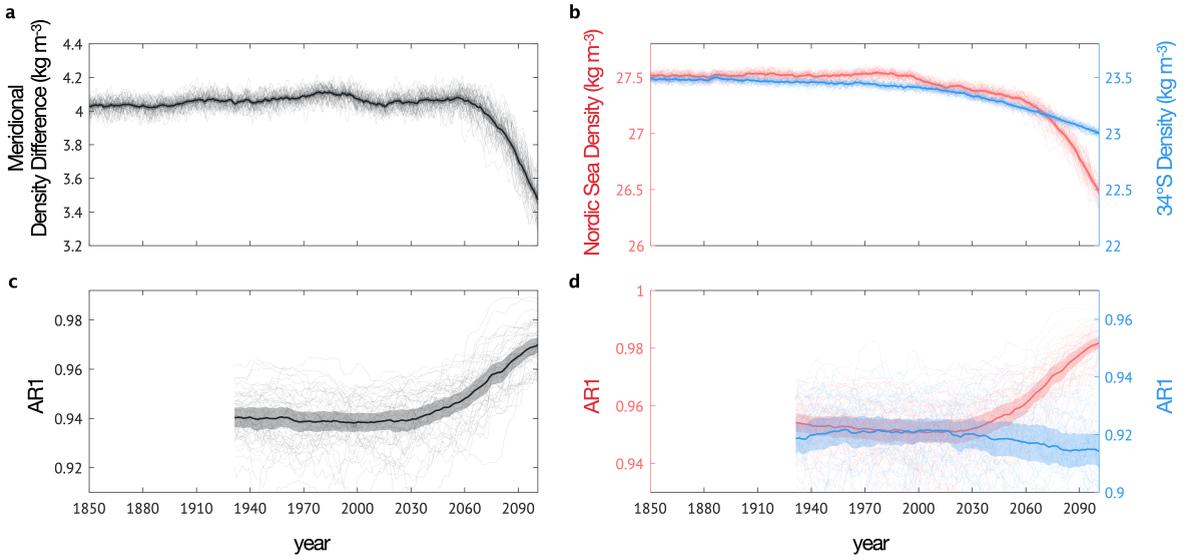

**Supplementary Fig. 8 Meridional density difference and CSD. a-d**, Time series of meridional density difference (**a**), northern boundary (Nordic Sea; red) and southern boundary (34°S; blue) (**b**), and corresponding lag-1 autocorrelation (AR1) (**c-d**). The thick line shows the ensemble mean, shading represents the 95% confidence interval, and thin lines indicate individual members.

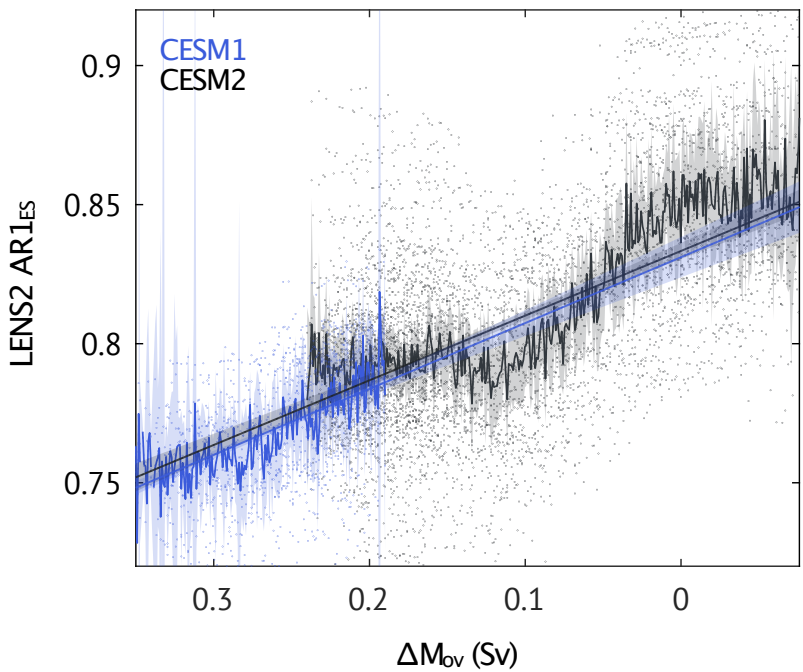

**Supplementary Fig. 9 Same as Fig. 4a, but with CESM1.** Eastern SPNA SST AR1 as a function of $\Delta M_{OV}$ for CESM2 (black) and CESM1 (blue). The solid line represents the AR1 mean, while shading indicates the 95% confidence interval of AR1. The fitting lines are obtained by the least square method, with the 95% confidence interval of linear regression indicated by shading. CESM1 includes 28 ensemble members comprising a 70-year control simulation at a fixed $CO_2$ concentration of 367 ppm, followed by a 140-year global warming experiment with $CO_2$ increasing at a rate of 1% per year.

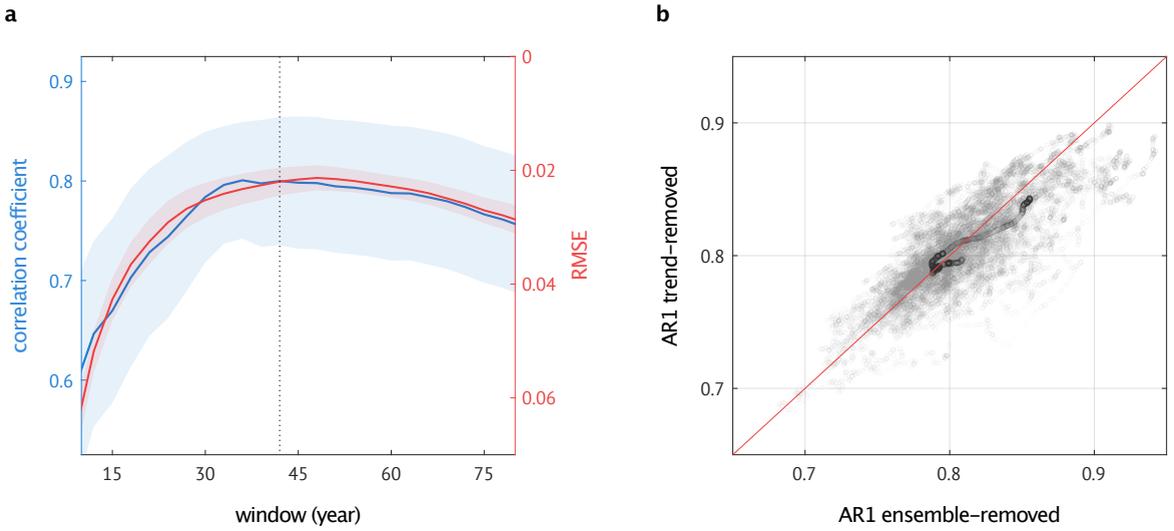

**Supplementary Fig. 10 Comparison of methodologies for deriving AR1 estimates. a**, (a) Correlation coefficients (blue) and root-mean square errors (RMSE; red) between AR1 values obtained from ensemble mean–removed residual time series and those derived from nonlinear trend–removed residual time series. Based on these metrics, a 44-year window is identified as optimal for reproducing the ensemble-mean forced response. **b,** Scatter plot comparing AR1 derived using ensemble-mean removal and those from the 44-year trend-removal approach. Grey circles represent individual ensemble members, black circles denote ensemble means, and the red line shows the one-to-one relationship.

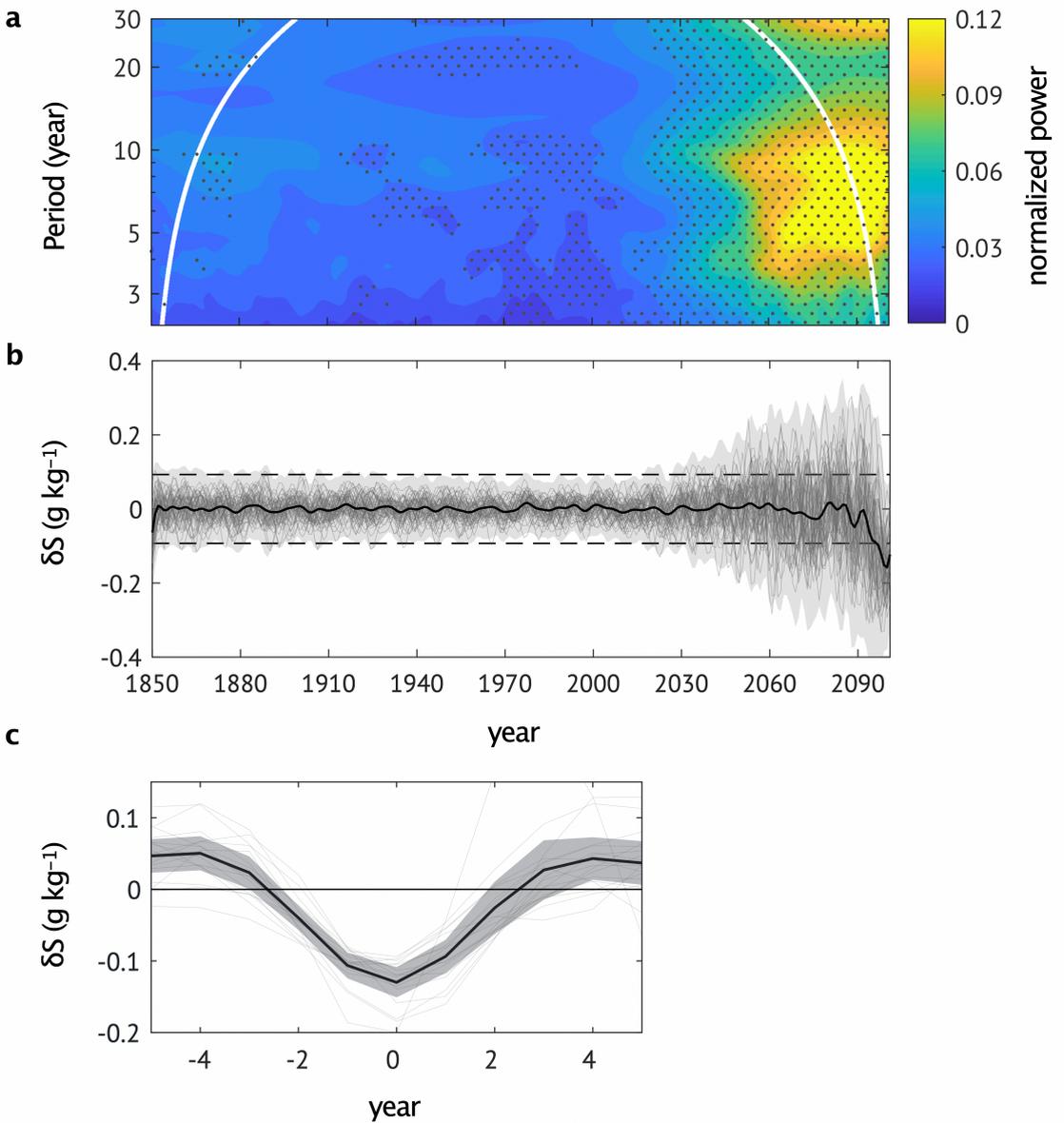

**Supplementary Fig. 11 Wavelet and composite of the eastern SPNA salinity. a**, Ensemble-averaged wavelet power spectrum of the upper-ocean (0-200 m) salinity in the eastern SPNA. The white line marks the cone of influence. Dotted areas indicate regions where the wavelet power is statistically significant at the 95% confidence level, as determined by bootstrap testing relative to the 1900–1950 baseline period. **b**, Time series of upper-ocean salinity anomalies, which is calculated 5-to-20 year bandpass filter. Gray lines represent individual ensemble members, and the black line indicates the ensemble mean. Light gray shading corresponds to ±3 standard deviations of salinity variability, with the horizontal dashed line showing the historical (1850–2000) ±3 standard deviation threshold. Dark gray shading denotes ±1 standard deviation. **c**, Composite time series derived by identifying freshening peaks when eastern SPNA salinity anomalies exceed the historical 3 standard deviation threshold during 2030–2060. Shading represents the 95% significance range.